\documentclass[journal]{IEEEtran}

\usepackage[hidelinks]{hyperref}
\hypersetup{
  colorlinks   = true, 
  urlcolor     = blue,
  linkcolor    = blue, 
  citecolor    = blue 
}

\usepackage{amssymb}
\usepackage{bm}
\usepackage{mathtools}
\usepackage{nccmath}
\usepackage{amsmath}

\usepackage{graphicx}
\graphicspath{{./images/}}
\DeclareGraphicsExtensions{.pdf,.png,.jpg}
\usepackage{placeins}
\usepackage[font+=scriptsize]{subcaption}
\usepackage{float}

\usepackage{booktabs}
\usepackage[table,xcdraw]{xcolor}
\usepackage{graphicx}
\usepackage{adjustbox}

\usepackage{lineno}
\usepackage{lipsum}
\usepackage{multicol}

\begin{document}

\title{Deep Fusion of Local and Non-Local Features for Precision Landslide Recognition}
\author{
    Qing Zhu, Lin Chen, Han Hu, Binzhi Xu, Yeting Zhang, Haifeng Li
    \thanks{Manuscript received: \today. This work was supported in part by the National Natural Science Foundation of China (Projects No.: 41941019, 41871291, 41871314) and in part by the National Key Research and Development Program of China (Project No.: 2018YFB0505404). (\emph{Mutual corresponding authors: Han Hu and Haifeng Li)}}
    \thanks{Qing Zhu, Lin Chen and Binzhi Xu are with the Faculty of Geoscience and Environmental Engineering, Southwest Jiaotong University, Chengdu, China. (e-mail: zhuq66@263.net; chenlin@my.swjtu.edu.cn; xubinzhi@my.swjtu.edu.cn)}
    \thanks{Han Hu is with the State Key Laboratory of Rail Transit Engineering Informatization, China Railway First Survey and Design Institute Co. Ltd., Xi'an, China and also with the Faculty of Geoscience and Environmental Engineering, Southwest Jiaotong University, Chengdu, China. (email: han.hu@swjtu.edu.cn)}
    \thanks{Yeting Zhang is with the State Key Laboratory of Information Engineering in Surveying, Mapping and Remote Sensing, Wuhan University, Wuhan, China (e-mail: zhangyeting@whu.edu.cn)}
    \thanks{Haifeng Li is with the School of Geosciences and Info-Physics, Central South University, Changsha, China (e-mail: lihaifeng@csu.edu.cn)}
}

\markboth{Preprint submitted to IEEE Geoscience and Remote Sensing Letters}{Xu \emph{et al.}, Deep Fusion of Local and Non-Local features for Precision Landslide Mapping}

\maketitle

\begin{abstract}
   Precision mapping of landslide inventory is crucial for hazard mitigation. Most landslides generally co-exist with other confusing geological features, and the presence of such areas can only be inferred unambiguously at a large scale. In addition, local information is also important for the preservation of object boundaries. Aiming to solve this problem, this paper proposes an effective approach to fuse both local and non-local features to surmount the contextual problem. Built upon the U-Net architecture that is widely adopted in the remote sensing community, we utilize two additional modules. The first one uses dilated convolution and the corresponding atrous spatial pyramid pooling, which enlarged the receptive field without sacrificing spatial resolution or increasing memory usage. The second uses a scale attention mechanism to guide the up-sampling of features from the coarse level by a learned weight map. In implementation, the computational overhead against the original U-Net was only a few convolutional layers. Experimental evaluations revealed that the proposed method outperformed state-of-the-art general-purpose semantic segmentation approaches. Furthermore, ablation studies have shown that the two models afforded extensive enhancements in landslide-recognition performance.
\end{abstract}


\IEEEpeerreviewmaketitle

\section{Introduction}
\label{s:1}

\IEEEPARstart{L}{andslide} is one of the most destructive natural hazards, and may also cause a series of secondary disasters, such as floods from barrier-lake overflows and dam breakages \cite{martha2011segment}.
Due to the complexity of factors that can cause landslides, and the abrupt occurrence of landslides during or after continuous rainfall in landslide-prone areas, pre-hazard mapping of landslide susceptibility is not sufficient for hazard management \cite{dou2015automatic}.
Efficient and precision mapping of post-hazard landslide regions is also crucial for successful emergency responses, and the prediction of secondary landslides that may occur due to unstable underlying surfaces.

Currently, many landslide-mapping systems require interactive interpretations, which obviously depend on the experience of human experts, and are thus not sufficient to enable rapid response.
Classical approaches tend to use domain-specific knowledge of the spectral characteristic of optical or radar images for the delineation and image-segmentation of landslides, such as textural patterns \cite{stumpf2011combining}, terrain features \cite{stumpf2011combining,rau2013semiautomatic} and vegetation indexes \cite{othman2013automatic}.
The object-based strategy \cite{blaschke2010object} is also widely used to increase the reliability when high-resolution data are used \cite{rau2013semiautomatic,othman2013automatic,keyport2018comparative,casagli2016landslide,dou2015automatic}.

With the advent of deep learning paradigms, convolutional neural network (CNN)-based approaches have yielded impressive results in many image-processing objectives.
Specifically, fully convolutional networks (FCNs) \cite{long2015fully} have enabled end-to-end segmentation using a deconvolution module.
FCNs and their successors \cite{chen2017deeplab, chen2018encoder} have significantly boosted the development of semantic segmentation of images.
However, some challenges remain to be surmounted before such techniques can be applied to the mapping of landslides.
(1) \emph{Local receptive field.} CNN-based features only use information in local regions, but landslides often have confusing spectral information generated by background features, such as roads and residential areas, and large contexts are required to remove these ambiguities during segmentation.
(2) \emph{Boundary preservation.} FCN essentially uses an up-sampling step to recover the low-resolution feature maps to the original resolution, which results in the loss of large amounts of boundary information.
Although pyramid structures such as Deeplab and U-Net \cite{chen2017deeplab,ronneberger2015u} can propagate and aggregate information from different scales, this problem persists.
The strategy of change detection has also been considered \cite{lei2019landslide} for better boundary preservation, but it is not applicable when no pre-hazard inventory exists.

In this study, we have proposed and developed an approach to alleviate the above problems, using the fusion of both local and non-local information, built upon the U-Net structure \cite{ronneberger2015u}.
During the down-sampling, we used the atrous convolution \cite{chen2017rethinking,chen2018encoder} with different dilation sizes to simulate multiscale features, prior to down-sampling via a max-pooling operation\cite{ronneberger2015u}. In addition, inspired by the Deeplab, we added a fusion step in the bottleneck of U-Net to aggregate multiscale features \cite{chen2017deeplab}.
In the up-sampling step, inspired by the scale attention module \cite{chen2016attention,oktay2018attention}, specifically, the alignment model \cite{jetley2018learn,xu2015show}, we augmented the U-Net with the attention module to suppress the irrelevant and confusing features from coarse scales by learning a weight map.
These two incremental modifications cooperated to improve the accuracy and robustness of the mapping of landslide regions using only post-hazard unmanned aerial vehicle (UAV) images.

\section{Methodology}
\label{s:method}

\subsection{Problem setup and overview of the approaches}
We formulated the mapping of the post-hazard landslide from UAV images as a semantic segmentation problem \cite{long2015fully} that is widely studied in the computer vision community.
More specifically, given an image $ \textbf{X} $, the purpose was to assign a binary label $ l_i~(l_i=\{0,1\}) $ to each pixel $ i $, which comprised the binary segmentation $ b $ of the image to background and landslide regions.
The objective was to learn this mapping $ \mathcal{F} \mapsto b $ in an end-to-end manner \cite{long2015fully}, using the training binary segmentation samples and corresponding UAV images.

\begin{figure}[h]
  \centering
  \includegraphics[width=\linewidth]{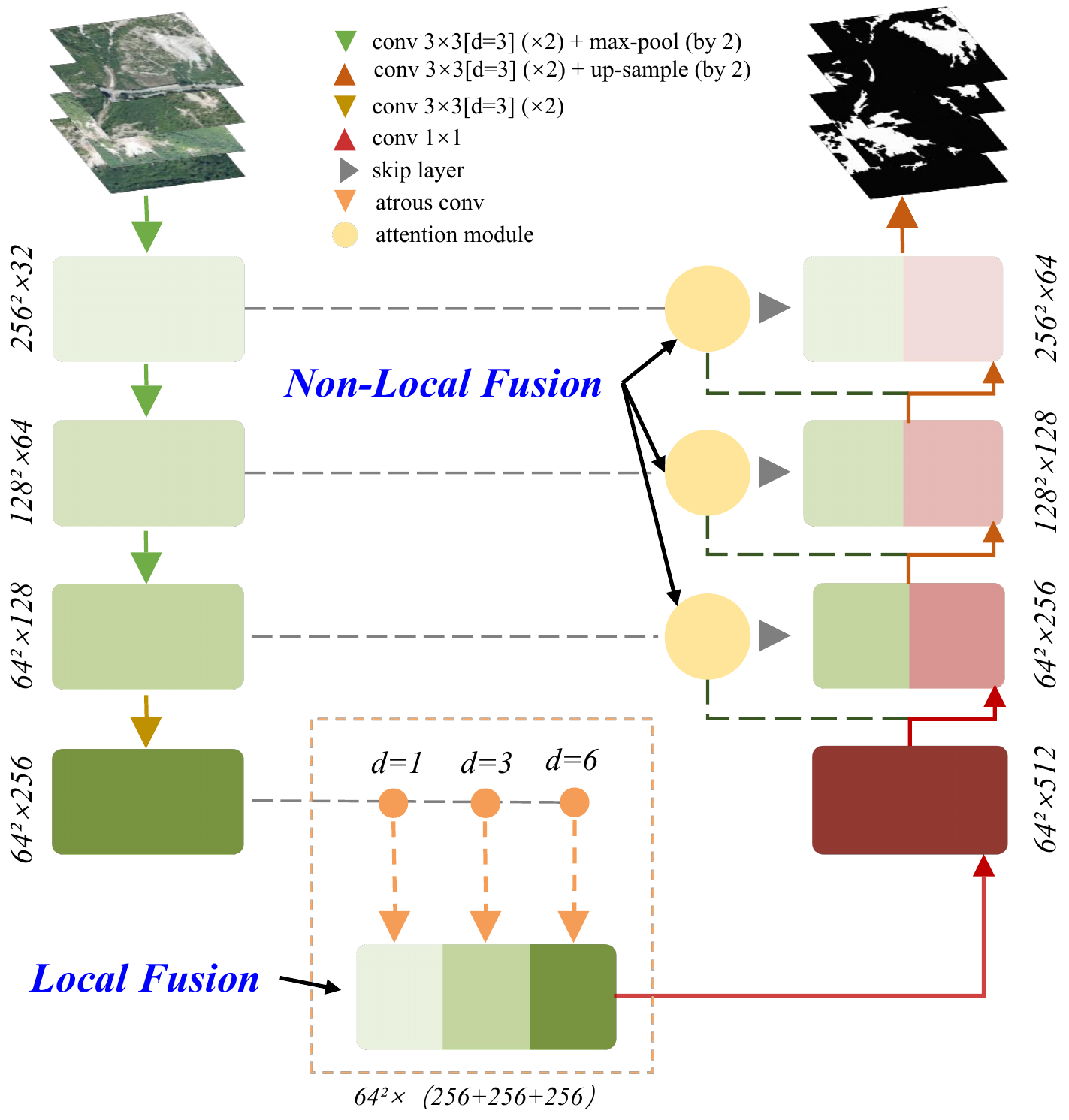}
  \caption{Overview of the architecture for the landslide mapping.}
  \label{fig:overview}
\end{figure}

Inspired by previous work on semantic segmentation of remote sensing images \cite{zhang2018road}, we built the system upon the prominent U-Net \cite{ronneberger2015u} structure, which fuses CNN maps in a multiscale fashion.
U-Net also features a low memory profile, which is critical for large-scale remote sensing applications.
In addition, U-Net can make dense semantic predictions by using the encoder-decoder strategy.
In the encoder, U-Net grasped both the low-level greyscale and gradient features and high-level contextual features from finer space resolution (shallower channels) to coarser space resolution (deeper channels, respectively).
In the decoder, U-Net concatenated the encoder features in the left part to the deconvolved features (Fig. \ref{fig:overview}). Its use of the above skip layers enabled fusion of multiscale information, which substantially improved the mapping resolution.

However, in our mapping of post-hazard landslide regions, we have often encountered regions with both very small and very large structures.
This presents a problem, as due to the computational efficiency, it is not possible to continuously increase the number of layers to enlarge the receptive field to encompass larger objects. In addition, the spectral signatures of small or local features may be obscured by areas with similar spectral information, such as bare earth or roads.

To overcome this barrier, in this study we augmented the U-Net with two modules: (1) the ASPP (atrous spatial pyramid pooling) module for enlarging the receptive field without going too deep and losing spatial resolution in the encoder; and (2) the attention module, to suppress irrelevant or confusing features by exploiting non-local contextual information at the coarse level in the decoder. The overview of the architecture is illustrated in Fig. \ref{fig:overview}. The two modifications are detailed in the subsections below.

\subsection{Atrous spatial pyramid pooling for enlarged receptive fields}
\subsubsection{Atrous convolution} Classical convolution is intrinsically a local method, which can only account for a fixed region and relies on pooling operations, e.g. max-pooling, to enlarge the receptive field at the cost of coarser spatial resolution. Atrous convolution \cite{chen2017deeplab,chen2017rethinking} has been used to surmount this problem, as it adds holes in the convolution kernel, which can effectively enlarge the receptive field without sacrificing spatial resolution or efficiency, as opposed to enlarging the size of the convolution kernel. The number of holes is controlled by the dilation rate $ d $, and the size of the kernel with holes $ k_h $ is also related to the effective kernel size $ k $ as $ k_h = k + (k-1)\times(d-1) $. Another strength of the atrous convolution approach is that dense feature maps are produced by sliding the window across the whole image. This property substantially simplifies the fusion of multiscale information.

\subsubsection{Atrous spatial pyramid pooling} 

Although atrous convolution can solve the receptive field problem, if the dilation rate is sequentially increasing across several dependent layers, the information will inevitably become too sparse \cite{chen2017rethinking}.
Therefore, we used an ASPP strategy \cite{chen2017deeplab} to substitute for the bottleneck part of the U-Net, as shown in Fig. \ref{fig:overview}.
Specifically, the fusion was denoted as 
\begin{equation}
  y=D_{3,1}(x) \oplus D_{3,3}(x) \oplus D_{3,6}(x),
  \label{eq:aspp}
\end{equation}
where $ D_{d,k} $ indicates the atrous convolution with a dilation rate of $ d $ and a kernel size of $ k $ and $ \oplus $ denotes the concatenation operation of the features. The dimensions of the fused features were resized through convolution with a $ 1\times1 $ kernel. By use of the ASPP fusion, the receptive field could account for more than a quarter of the entire tile.


\subsection{Attention augmented up-sampling}

The attention mechanism originated from natural language processing \cite{vaswani2017attention}, and was later extended to image classification \cite{jetley2018learn} and semantic segmentation \cite{chen2016attention,wang2018non,oktay2018attention,fu2019dual}.
The premise of the attention mechanism is that the absolute values of feature maps also reveal their importance; therefore, we can learn a weight map $ \alpha \in (0,1) $, with the same spatial resolution as the feature map, to adaptively suppress irrelevant features.
Typically, such a weight map considers global information \cite{chen2016attention,fu2019dual} and two models (also known as the compatibility functions \cite{jetley2018learn}) are available, the additive model and the multiplicative model.
The success of the dual attention network \cite{fu2019dual} in general-purpose semantic segmentation motivated us to extend the additive model to the U-Net structure, i.e., by using the scale attention module introduced in previous works \cite{chen2016attention,oktay2018attention}.

Specifically, we did not directly concatenate the feature map from the encoder (\emph{i.e.} left part of Fig. \ref{fig:overview}) with the up-sampled feature map in the coarser level from the decoder (\emph{i.e.} right part of Fig. \ref{fig:overview}).
Instead, we first augmented the encoded features using the features from the coarser level of the decoder.
As features in the coarser level contained more contextual information, the non-local information helped in the determination of the weight map.

\begin{figure}[h]
  \centering
  \includegraphics[width=\linewidth]{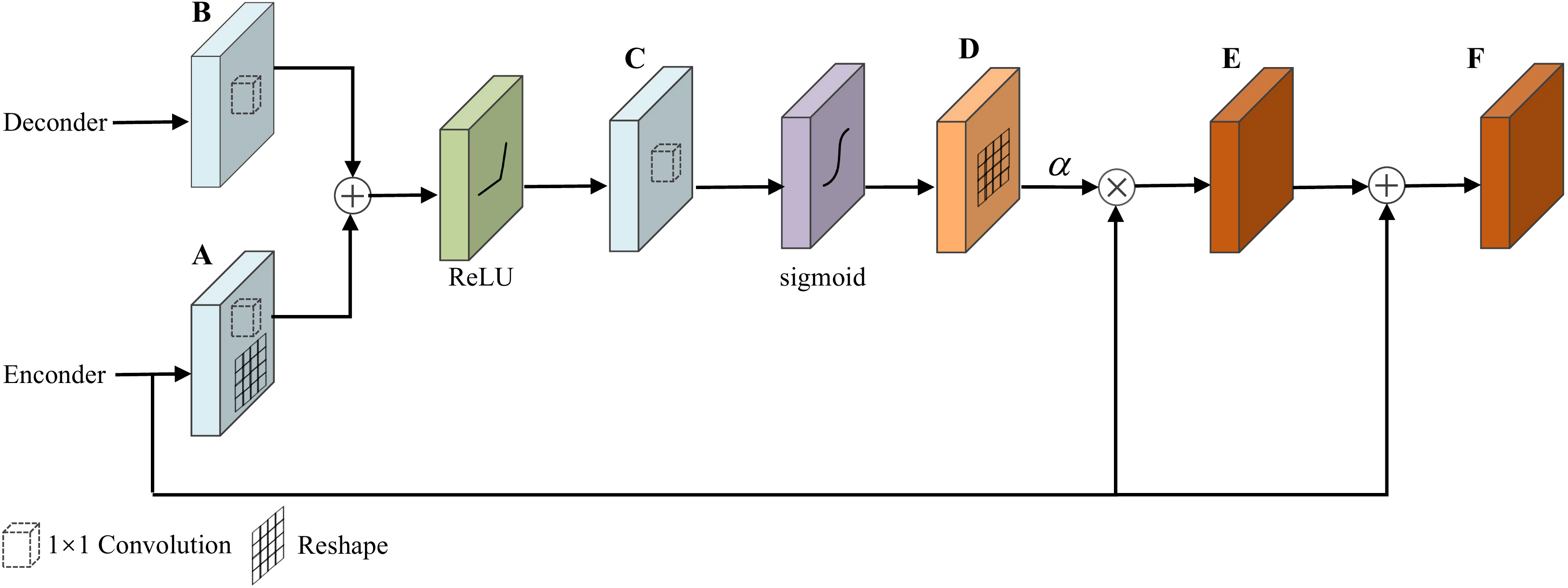}
  \caption{Attention augmented up-sampling by the non-local fusion of features from coarser levels.}
  \label{fig:attention}
\end{figure}

Fig. \ref{fig:attention} shows the enlarged representation of the attention module, \emph{i.e.} the circle in Fig. \ref{fig:overview}. The input encoder and decoder features, \emph{i.e.} $\textbf{A}$ and $ \textbf{B} $ respectively, were first convolved with a $ 1\times1 $ filter to make the channels compatible; in addition, the encoder features $ \textbf{A} $ were also reshaped to make the size compatible. The two maps were then connected by an element-wise summation, followed by processing via a rectified linear activation (ReLU) function. Then, another convolution with only $ 1 $ channel was used to create the compatibility map $ \textbf{C} $ \cite{jetley2018learn}.

Two options were available to create the attention map $ \alpha $: the \emph{softmax} $ \theta(x_i)=\tfrac{e^{x_i}}{\sum_t{e^{x_t}}} $ and \emph{sigmoid} $\theta(x_i)=\tfrac{e^{x_i}}{e^{x_i}+1}$ function. Although most approaches have used the \emph{softmax} function \cite{chen2016attention,jetley2018learn,fu2019dual}, we have found that the normalisation part in the \emph{softmax} function, \emph{e.g.} $\sum_i{\alpha_i}=1$, will make the activation too sparse, which is not good for a process that is applied multiple times. Therefore, the \emph{sigmoid} function was used to generate the feature $ \textbf{D} $ and resampled to the spatial resolution of the encoder feature $ A $, similar to previous work \cite{oktay2018attention}. The attention-augmented feature $\textbf{E}$ comprised element-wise multiplication with weight $ \alpha $. Finally, a skip connection \cite{he2016deep} was used for the map $ \textbf{F} $ before concatenation with the up-sampled features, as below,
\begin{equation}
  F_i = A_i + \alpha_iA_i.
\end{equation}

\subsection{Implementation details}
The labels were obtained interactively from the original UAV orthophotos in ArcMap and tiled into clips with a size of $ 512\times512 $.
The images were normalised to $[-0.5,0.5]$ for both training and testing.
As landslides are mainly located in forests, we performed hard mining by intentionally sampling more confusing regions, such as bare earth, large rocks, roads and rivers.
Random flipping, rotation and scaling were used for the data augmentation process.
Tensorflow 1.9 was used to implement the framework on a machine with four NVIDIA RTX Titan graphics computing units (GPUs).
For the hyper-parameters of the training, the batch size was 12 for each GPU, momentum was $0.9$, learning rate was $1e^{-3}$ and regularisation was $ 5e^{-4} $.
The training lasted for 100000 iterations and we recorded the model every 20000 iterations.
The model with the best testing performance was chosen. Finally, 
\emph{Softmax} was used, incorporating the binary segmentation loss function.
During testing, the original orthophotos were clipped, loaded and mosaicked dynamically.

\section{Experimental evaluations}

\subsection{Experimental setup and overall performance}
The UAV images covering six counties were obtained for the landslides caused by the earthquake in Jiuzhaigou, China on 8 August 2017.
We interactively selected $ 10^4 $ tiles for the training, and used a $ 70-30\% $ spilt. Four entire UAV orthophotos were used for the testing, which comprised approximately 3000 tiles. Three common metrics were used to evaluate the pixel-wise results, namely $precision$, $recall$ and $F_1$ score.

In the following, we use the prefix ``D'' to denote the dilated convolution and ``A'' to denote the attention module, such as D-U-Net augmented with only dilated convolution and DA-U-Net with both modified modules. For comparison, we also reimplemented several publicly available methods, such as the FCN \cite{long2015fully} with VGG-16 \cite{simonyan2014very} as backbone, PSPNet \cite{zhao2017pyramid} with ResNet-101 \cite{he2016deep} as backbone, the latest DeepLabV3+ \cite{chen2018encoder} with ResNet-101 as backbone and the vanilla U-Net \cite{ronneberger2015u}.

\begin{figure*}[ht]
  \centering
  \subcaptionbox{Orthophoto}[0.128\textwidth]{\includegraphics[width=0.128\textwidth]{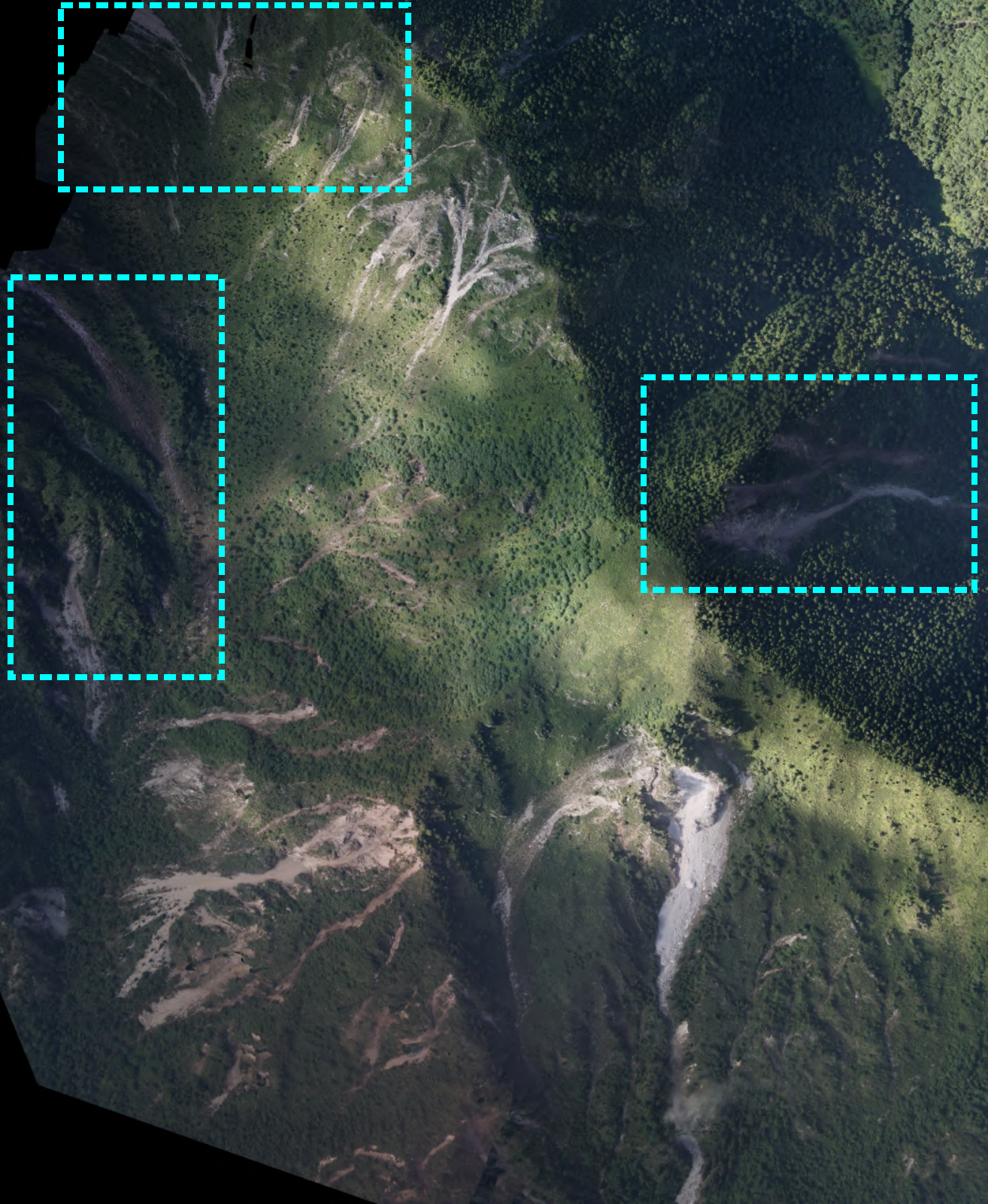}}
  \subcaptionbox{Label}[0.128\textwidth]{\includegraphics[width=0.128\textwidth]{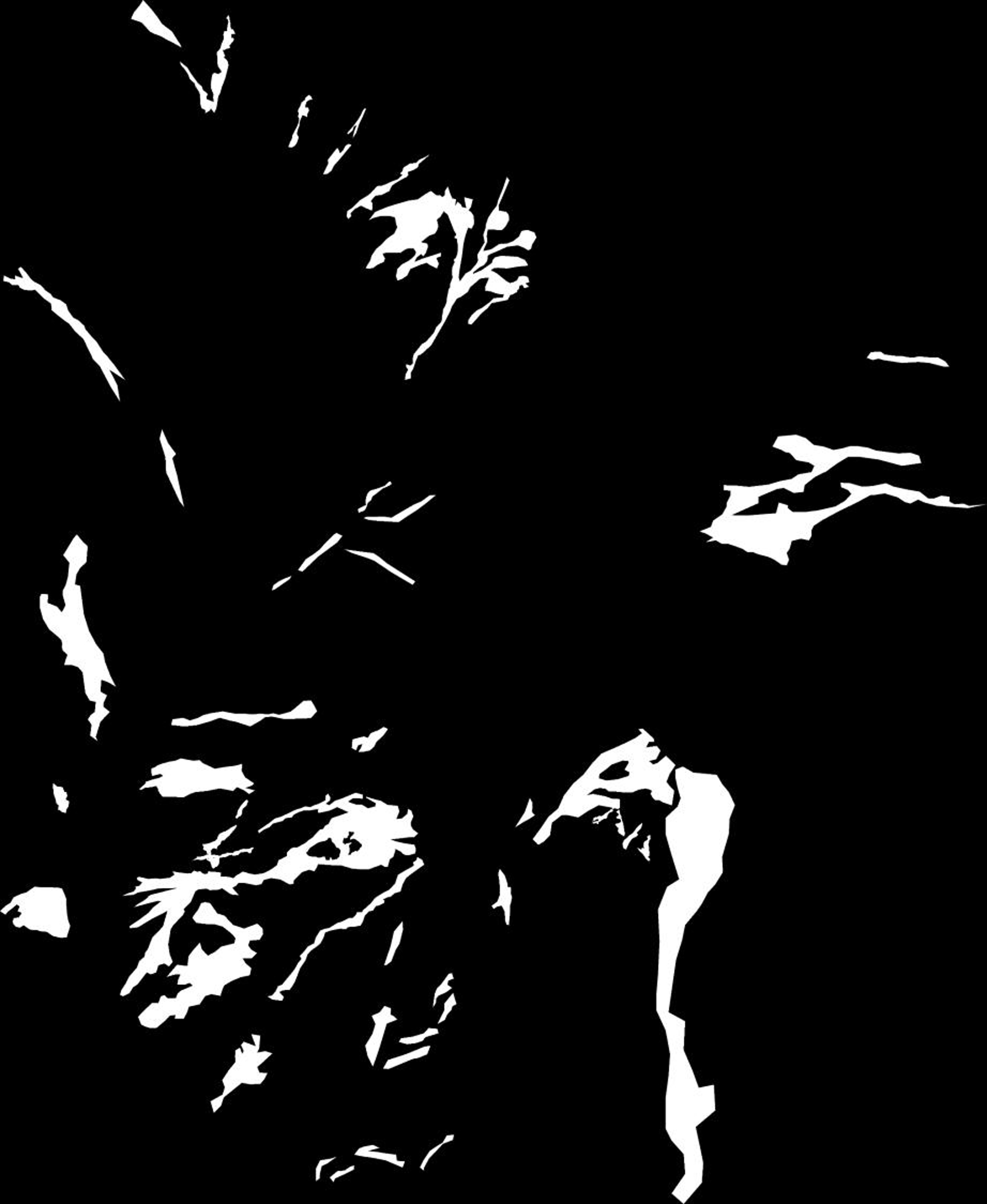}}
  \subcaptionbox{FCN}[0.128\textwidth]{\includegraphics[width=0.128\textwidth]{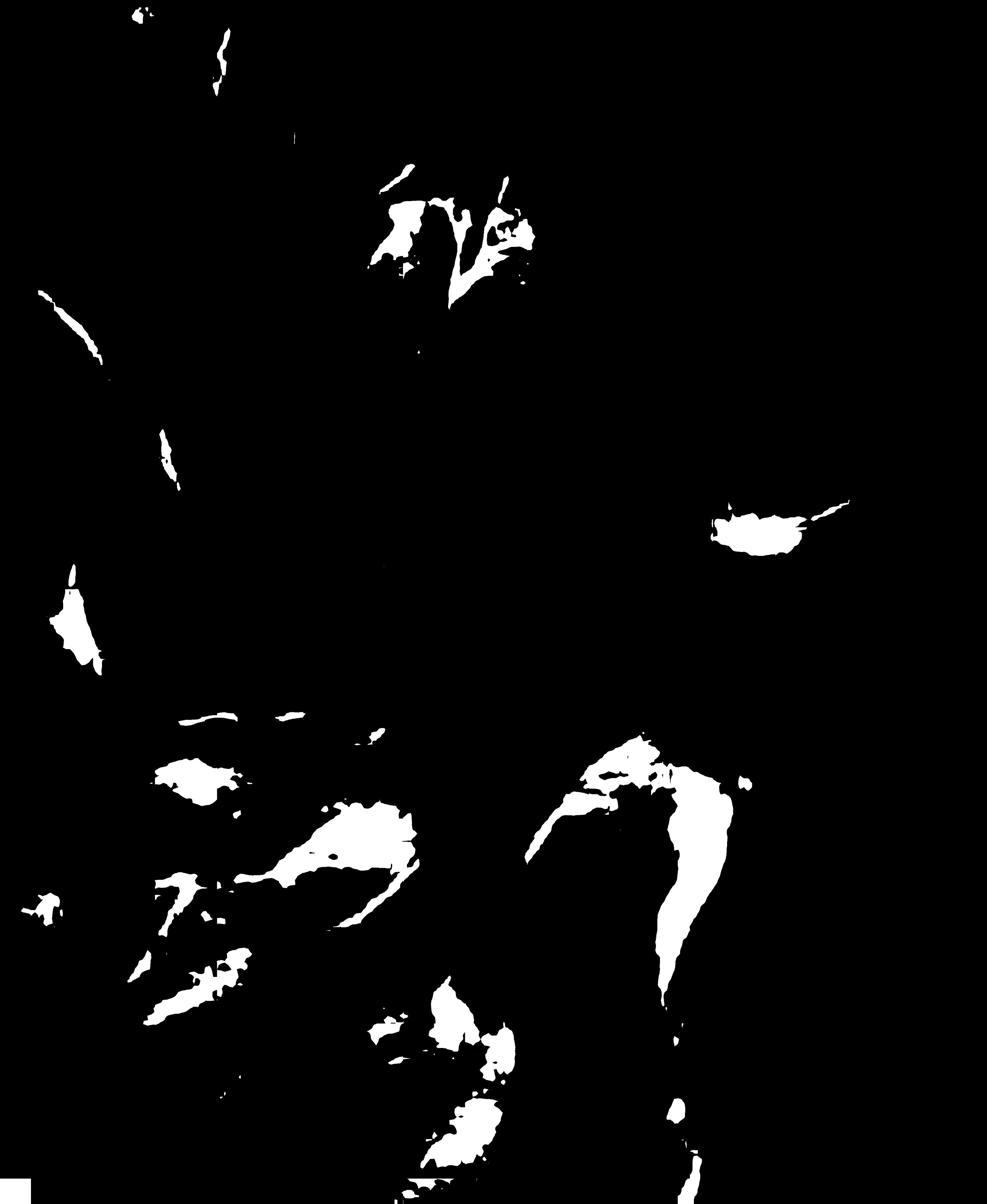}}
  \subcaptionbox{PSPNet}[0.128\textwidth]{\includegraphics[width=0.128\textwidth]{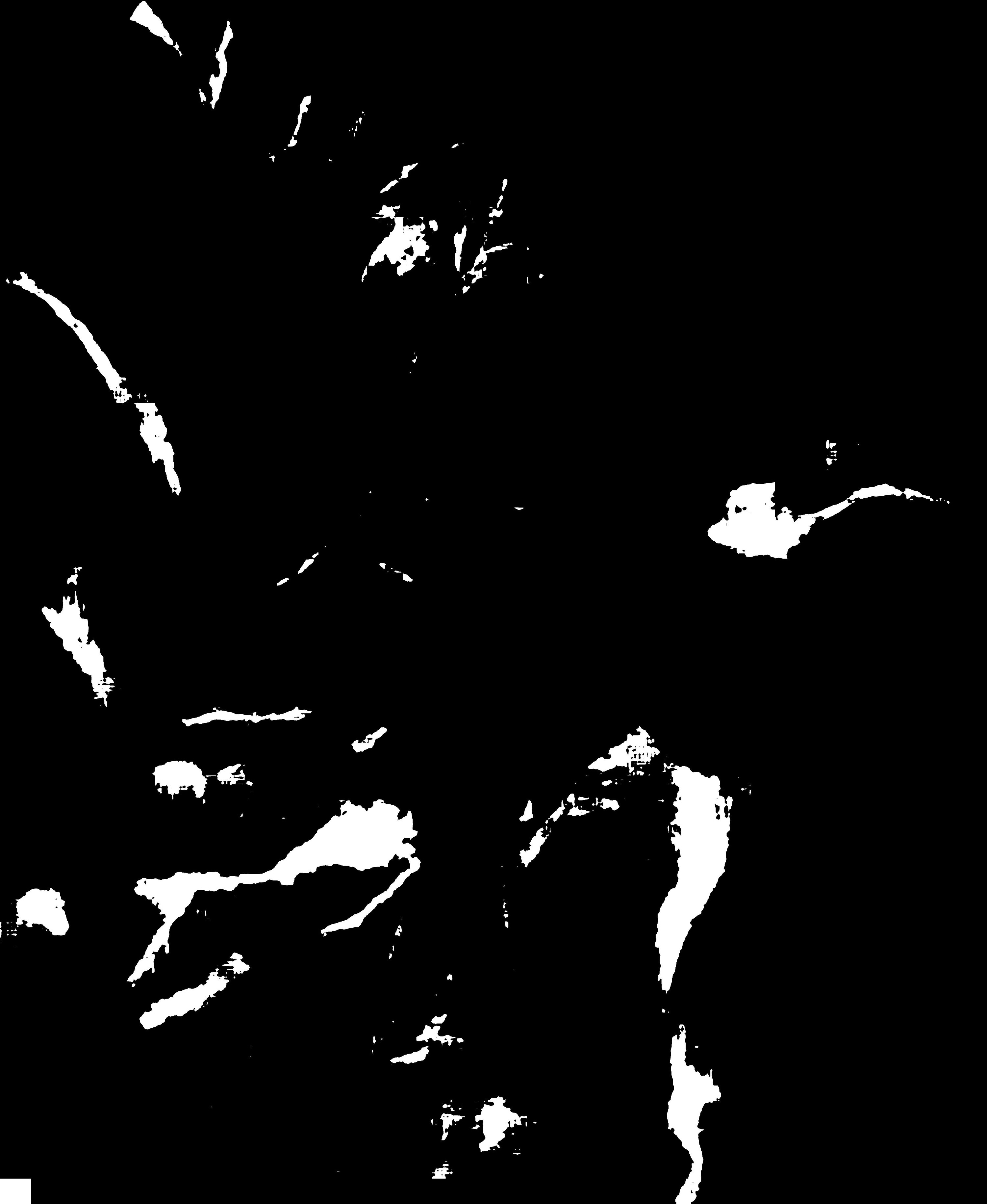}}
  \subcaptionbox{DeeplabV3+}[0.128\textwidth]{\includegraphics[width=0.128\textwidth]{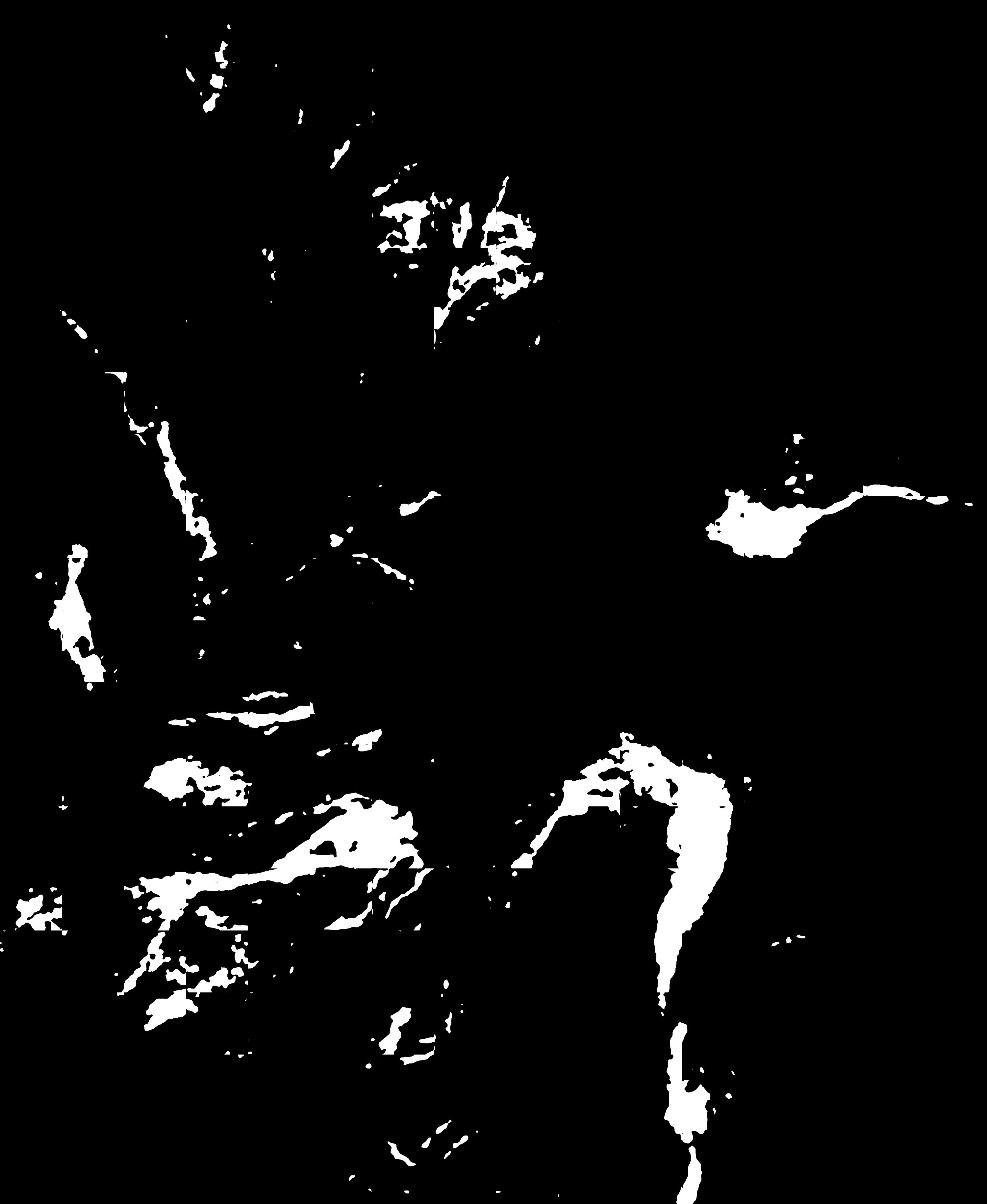}}
  \subcaptionbox{U-Net}[0.128\textwidth]{\includegraphics[width=0.128\textwidth]{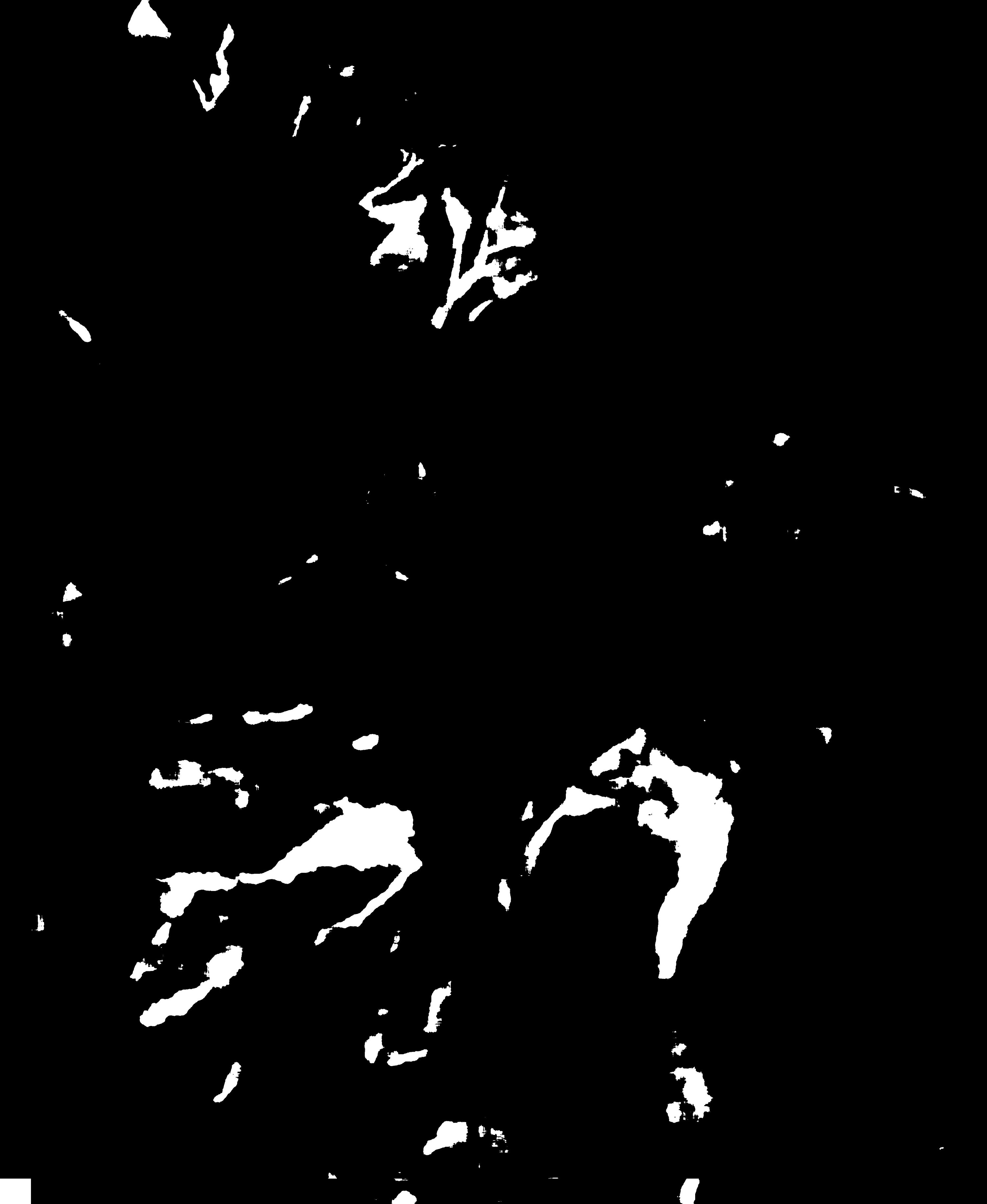}}
  \subcaptionbox{DA-U-Net}[0.128\textwidth]{\includegraphics[width=0.128\textwidth]{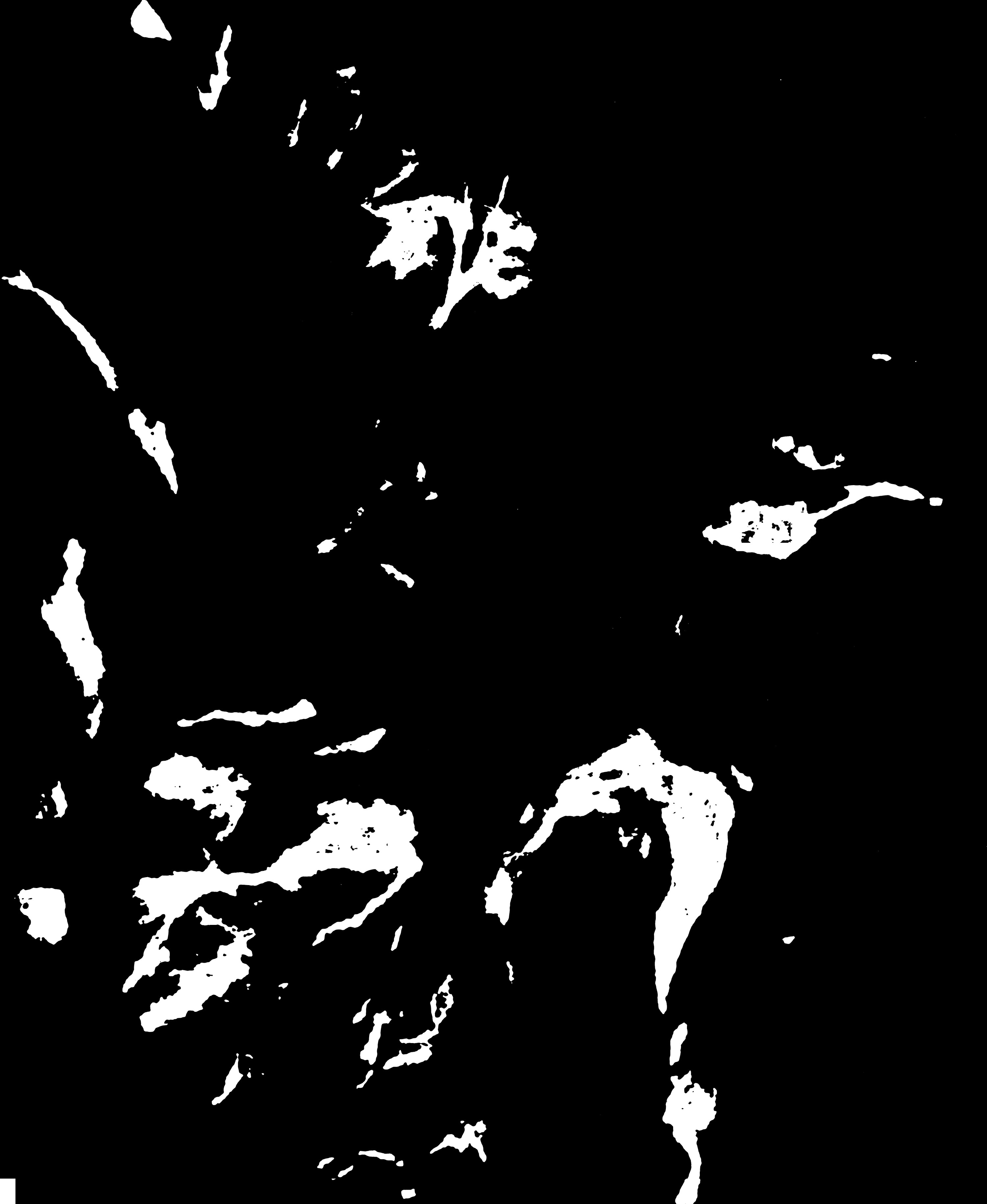}}
  \caption{Qualitative comparisons with other methods.}
  \label{fig:qualitative}
\end{figure*}

Fig. \ref{fig:qualitative} compares the performances of the methods above in assessing a typical scene. Notably, in the shadow (eclipse region), almost all of the methods fail to identify the landslide region; this situation could only be improved with use of the attention module, e.g., DA-U-Net. Another interesting finding is that only architectures with a pyramid strategy can satisfactorily identify landslide regions in confusing areas comprising both landslides and bare earth (as indicated by the rectangle). In summary, the proposed methods gave the best overall segmentation results. 

Turning to quantitative evaluations, Table \ref{tab:quantitative} demonstrates the Intersection of Union (IoU), precision, recall and F-scores for all of the methods. The proposed DA-U-Net exceeds the performance of the second-best method, \emph{i.e.} DeeplabV3+ \cite{chen2018encoder}, in the most concerned IoU metric and also for recall rate and F-score. Considering that the performance of the current state-of-the-art DeeplabV3+ is also on par with the PSPNet, the modifications on the U-Net were thus effective.

\begin{table}[h]
    \centering
    \caption{Quantitative comparisons of different methods. The bold cells denote the methods with best performances.}
    \label{tab:quantitative}
    \begin{tabular}{@{}lcccc@{}}
    \toprule
    Method     & IoU     & Precision & Recall & F-Score \\ \midrule
    FCN        & 48.15   & 75.96     & 56.81  & 65.00   \\
    U-Net      & 48.18   & 75.42     & 57.15  & 65.03   \\
    PSPNet     & 52.63   & 77.56     & 62.08  & 68.97   \\
    DeeplabV3+ & 57.25   & \textbf{79.6}      & 67.1   & 72.81   \\
    DA-U-Net   & \textbf{59.41}   & 70.06     & \textbf{79.62}  & \textbf{74.54}   \\ \bottomrule
    \end{tabular}
\end{table}

\subsection{Study of confusing areas}

\begin{figure*}[htb]
  \centering
  \subcaptionbox{Orthophoto}[0.1\textwidth]{
    \includegraphics[width=0.1\textwidth]{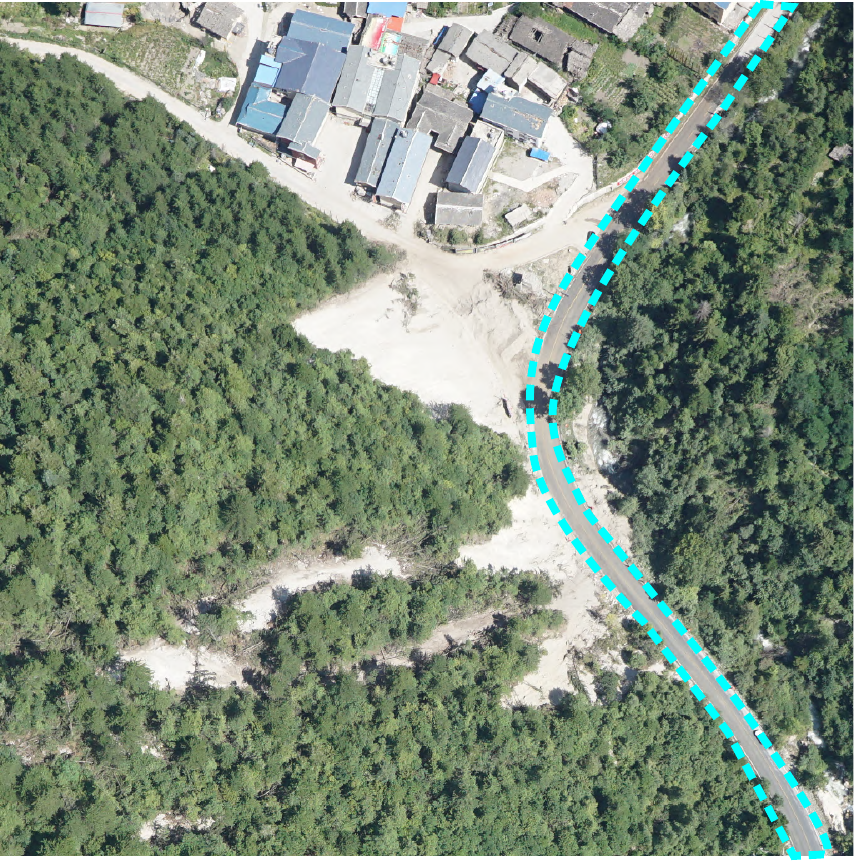}\vspace{0.3em}
    \includegraphics[width=0.1\textwidth]{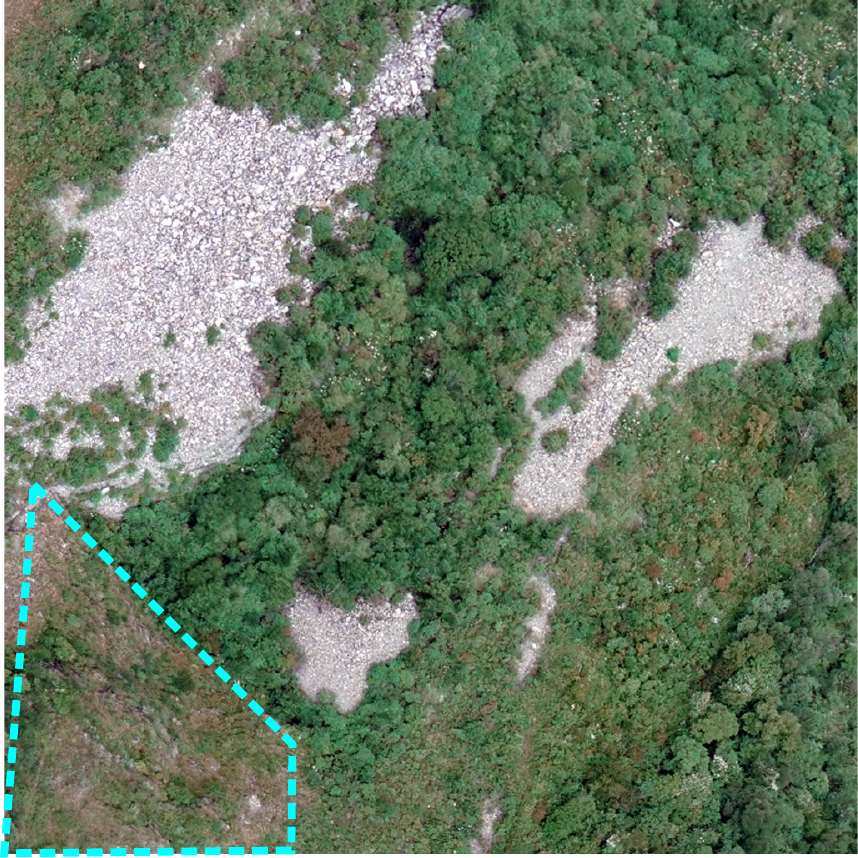}
  }
  \subcaptionbox{Label}[0.1\textwidth]{
    \includegraphics[width=0.1\textwidth]{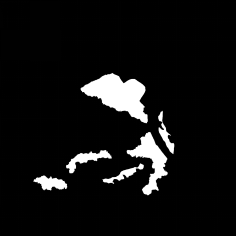}\vspace{0.3em}
    \includegraphics[width=0.1\textwidth]{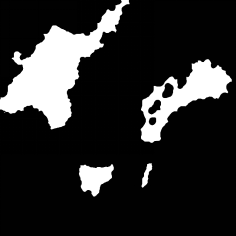}
  }
  \subcaptionbox{FCN}[0.1\textwidth]{
    \includegraphics[width=0.1\textwidth]{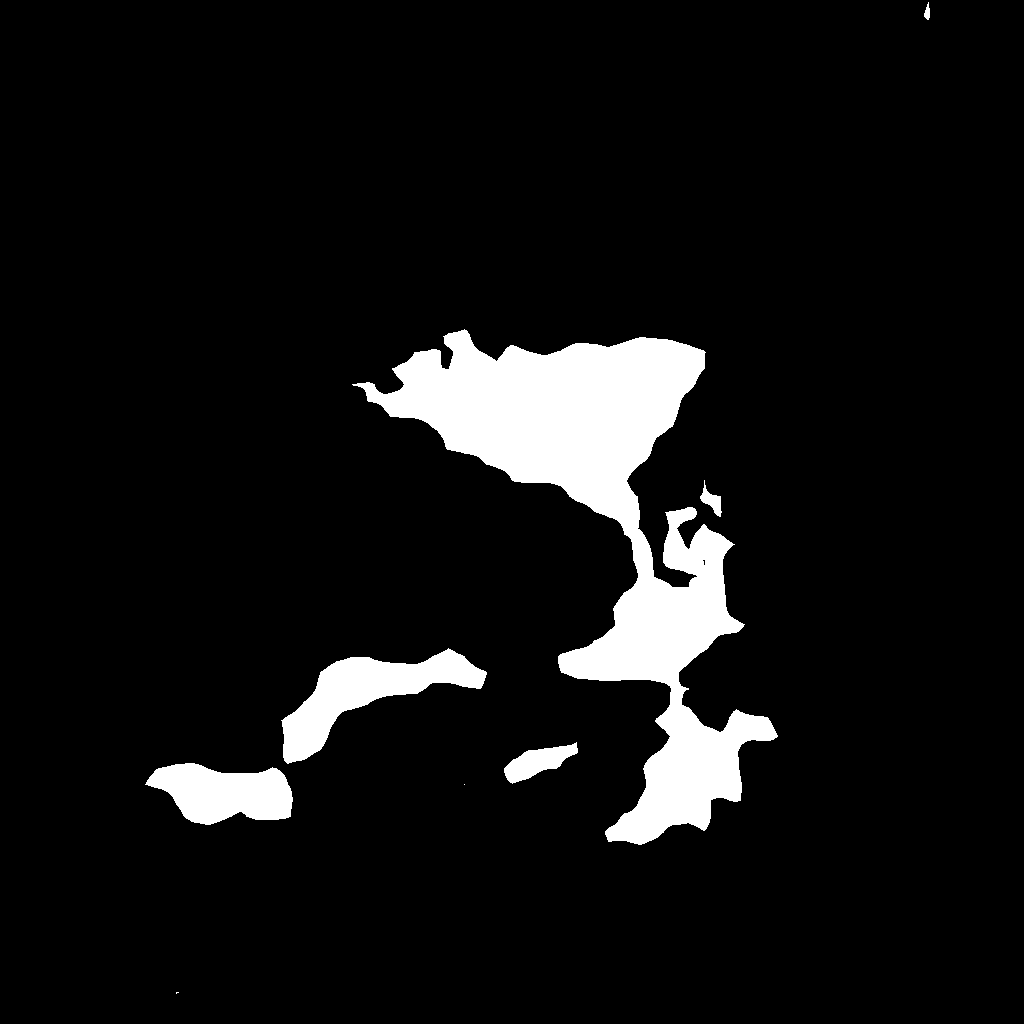}\vspace{0.3em}
    \includegraphics[width=0.1\textwidth]{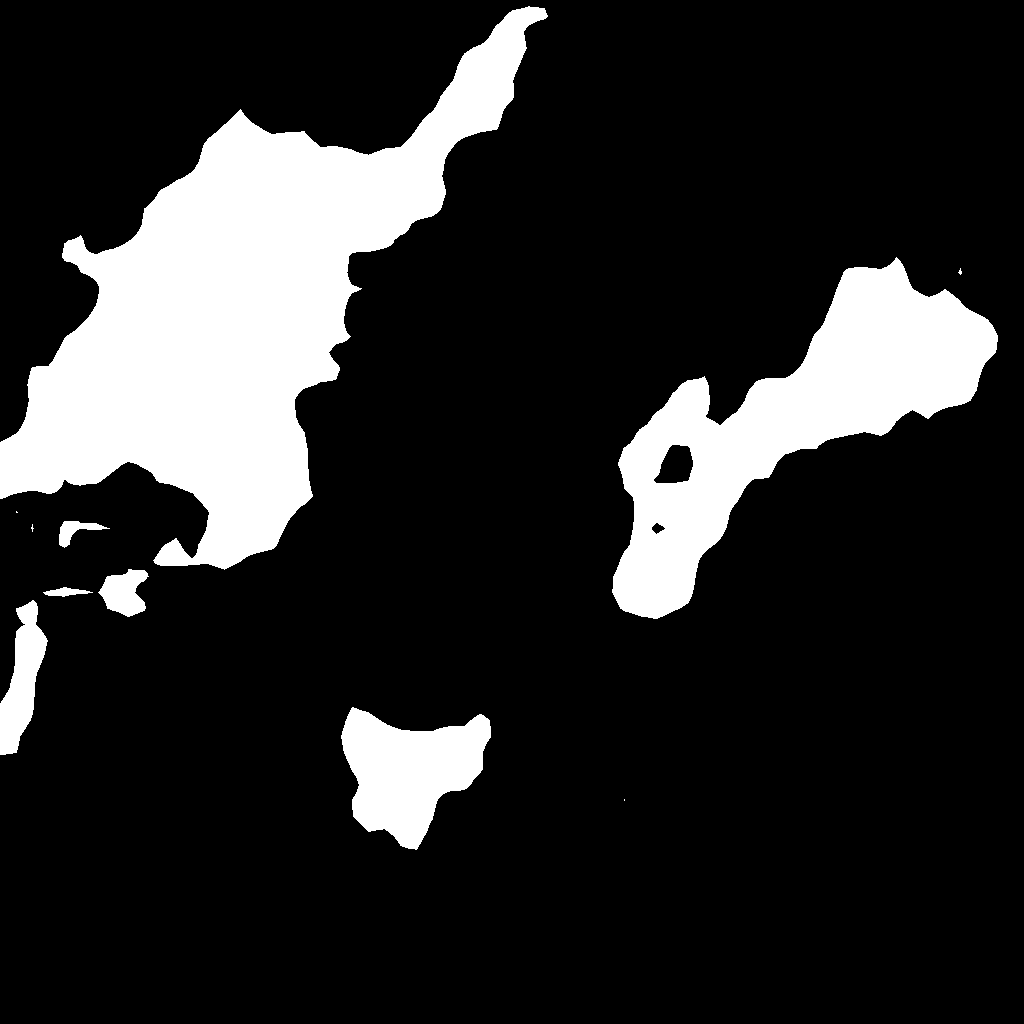}
  }
  \subcaptionbox{PSPNet}[0.1\textwidth]{
    \includegraphics[width=0.1\textwidth]{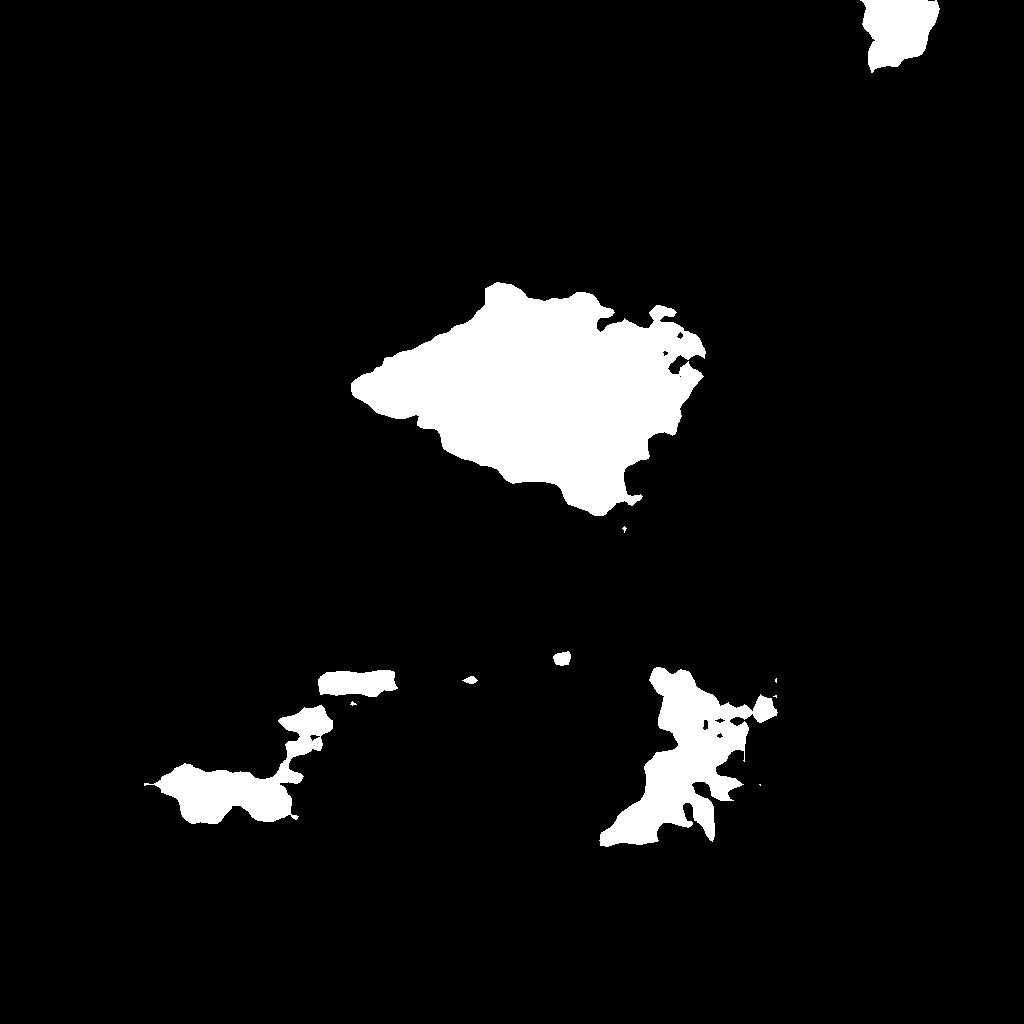}\vspace{0.3em}
    \includegraphics[width=0.1\textwidth]{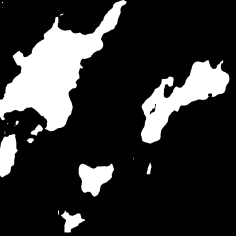}
  }
  \subcaptionbox{DeeplabV3+}[0.1\textwidth]{
    \includegraphics[width=0.1\textwidth]{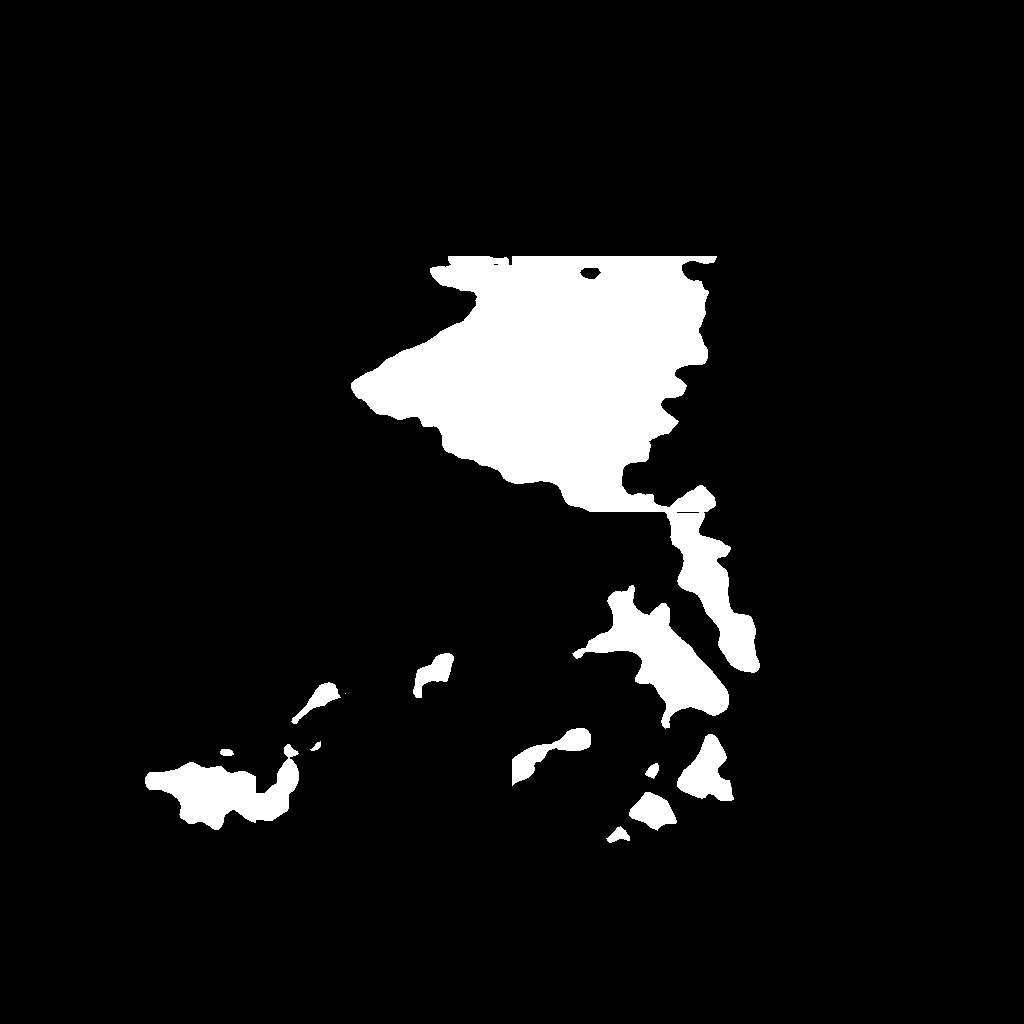}\vspace{0.3em}
    \includegraphics[width=0.1\textwidth]{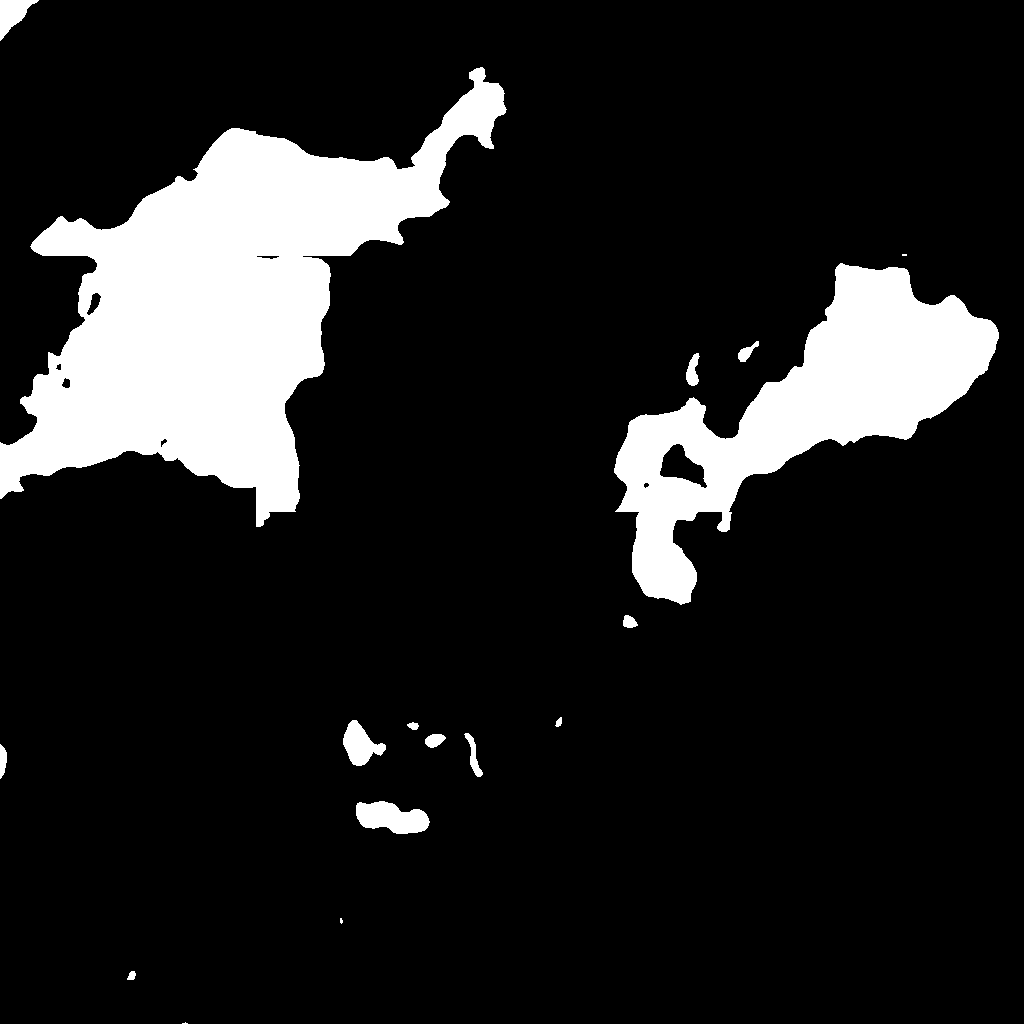}
  }
  \subcaptionbox{U-Net}[0.1\textwidth]{
    \includegraphics[width=0.1\textwidth]{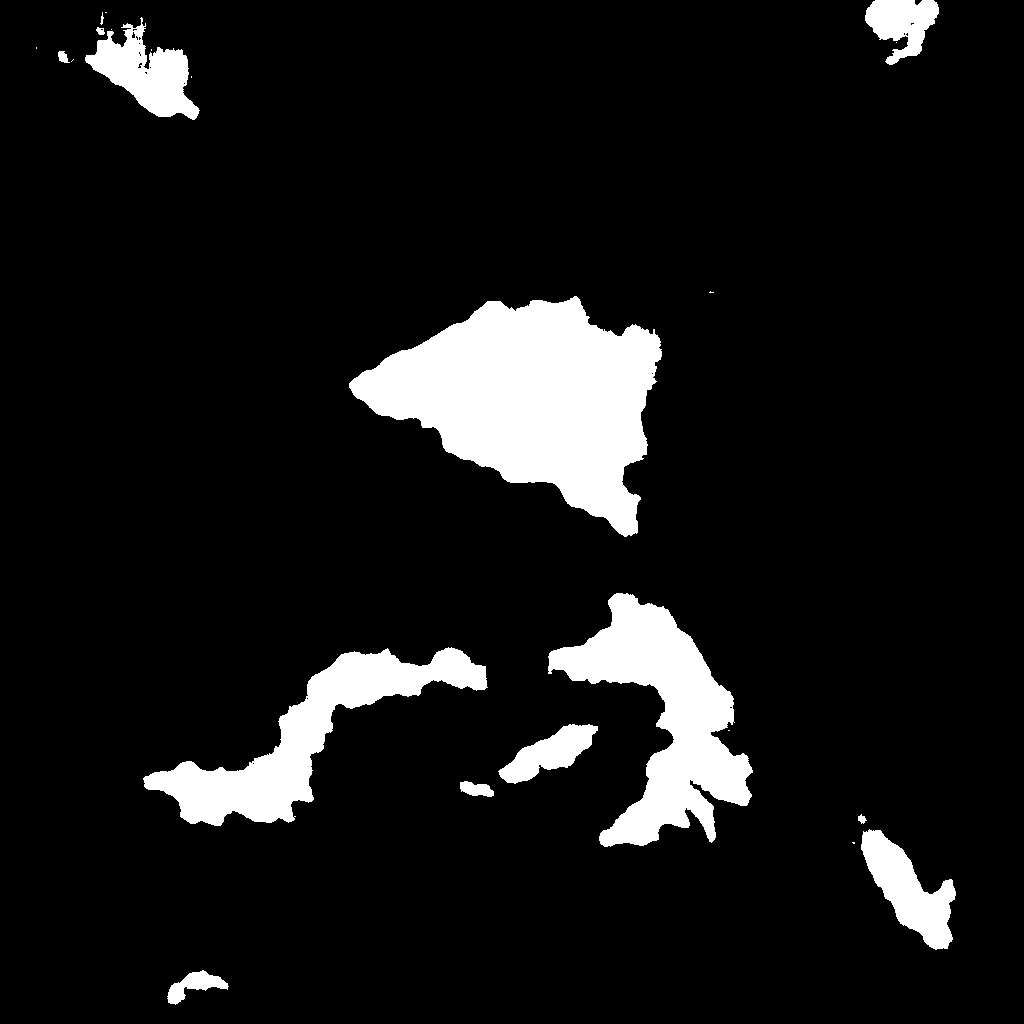}\vspace{0.3em}
    \includegraphics[width=0.1\textwidth]{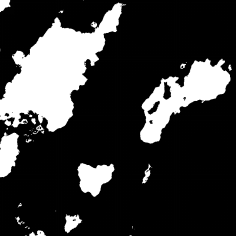}
  }
  \subcaptionbox{DA-U-Net}[0.1\textwidth]{
    \includegraphics[width=0.1\textwidth]{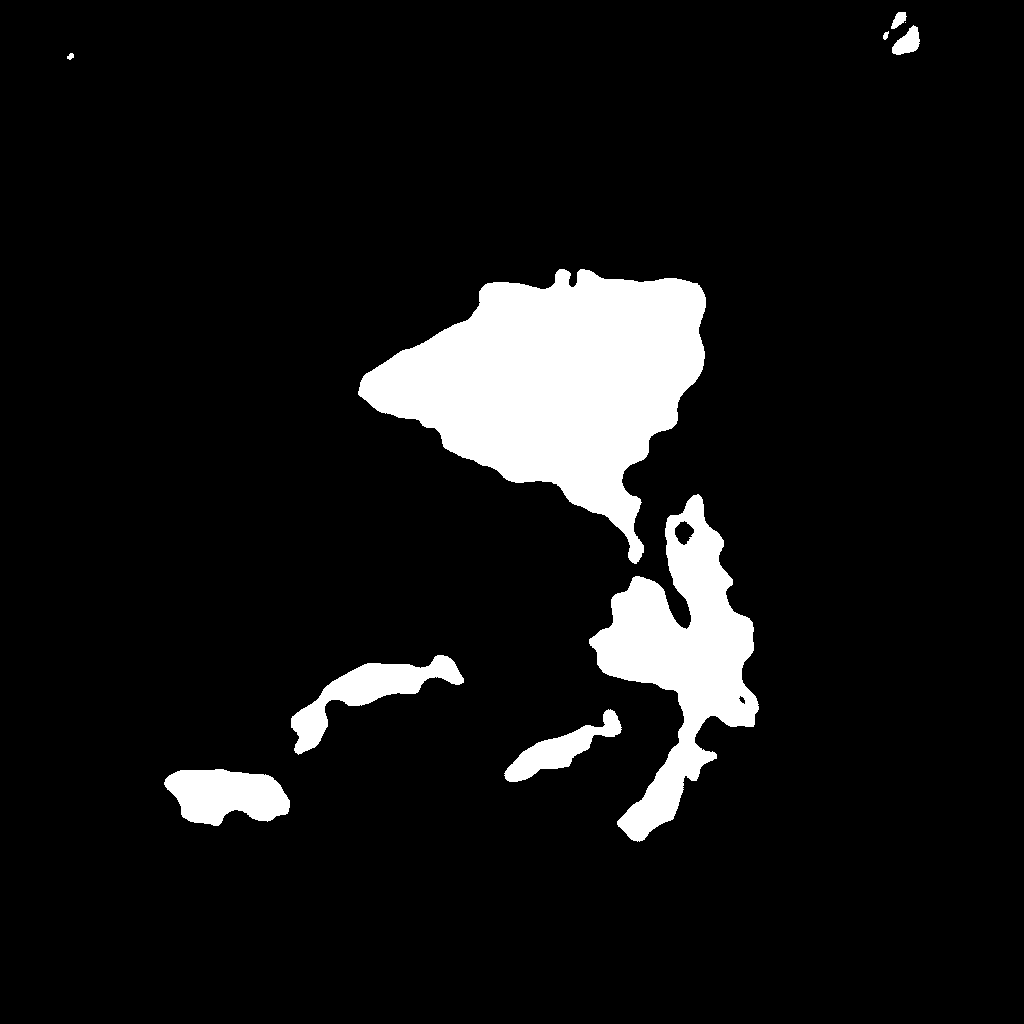}\vspace{0.3em}
    \includegraphics[width=0.1\textwidth]{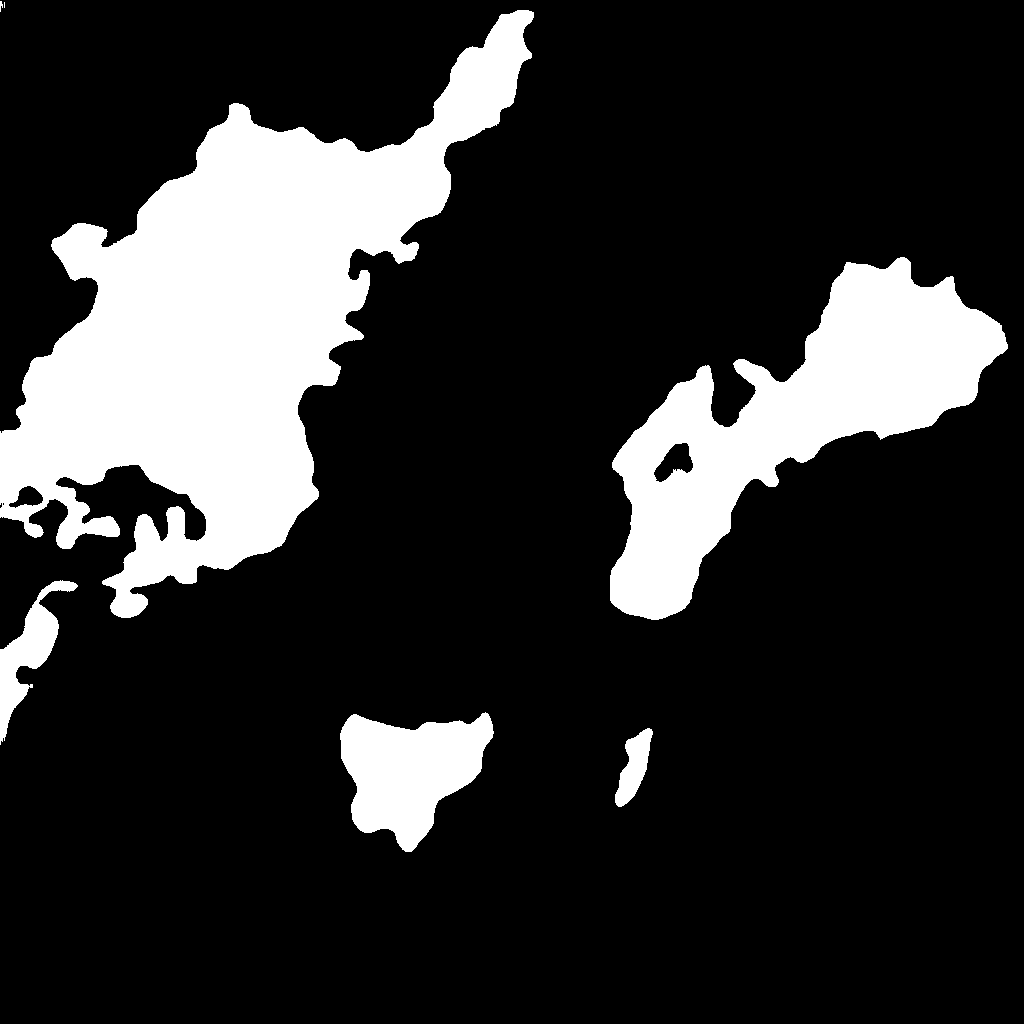}
  }
  \caption{Performances of different methods on typical confusing areas. The highlighted are common confusing regions.}
  \label{fig:confusing}
\end{figure*}

For efficient hazard mitigation and rapid emergency response, the most critical landslide regions that require precision detection are those that occur in confusing areas. As most landslide regions in the training were in non-confusing regions, we privileged confusing areas by choosing imbalanced samples that preferred the confusing regions. Fig. \ref{fig:confusing} shows two typical confusing areas, namely roads (top) and bare earth (bottom), with the interesting regions in each row highlighted with cyan polygons. As these regions are hard to distinguish without inferring from a large context, it is almost impossible to identify these confusing areas without fusion of multiscale features. 
Unfortunately, FCNs have no mechanism of handling multiscale features and therefore would be expected to have inferior performance. U-Net propagates and aggregates limited multiscale features by down-sampling, which means that it cannot go too deep in the contextual information without loss of spatial resolution. PSPNet constructs the spatial pyramid and fuses different layers and DeeplabV3 embeds the ASPP module for multiscale features; both of these can exploit larger contextual information without loss of spatial resolution, but do not sufficiently preserve fine-grained structures. The proposed DA-U-Net has the best performances, thanks to its ASPP module and attention-guided up-sampling.

\subsection{Ablation studies}

The DA-U-Net augments the vanilla U-Net with two modules: 1) the dilated convolution and ASPP in the bottleneck for the exploitation of larger contextual information; and 2) the attention module for guided up-sampling. Table \ref{tab:ablation} compares different variants against the vanilla U-Net, namely D-U-Net, A-U-Net and DA-U-Net. Notably, it can be seen that both the dilation and attention modules clearly and extensively improve the overall performance.

\begin{figure*}[ht]
  \centering
  \subcaptionbox{Image}[0.1\linewidth]{
    \includegraphics[width=0.1\linewidth]{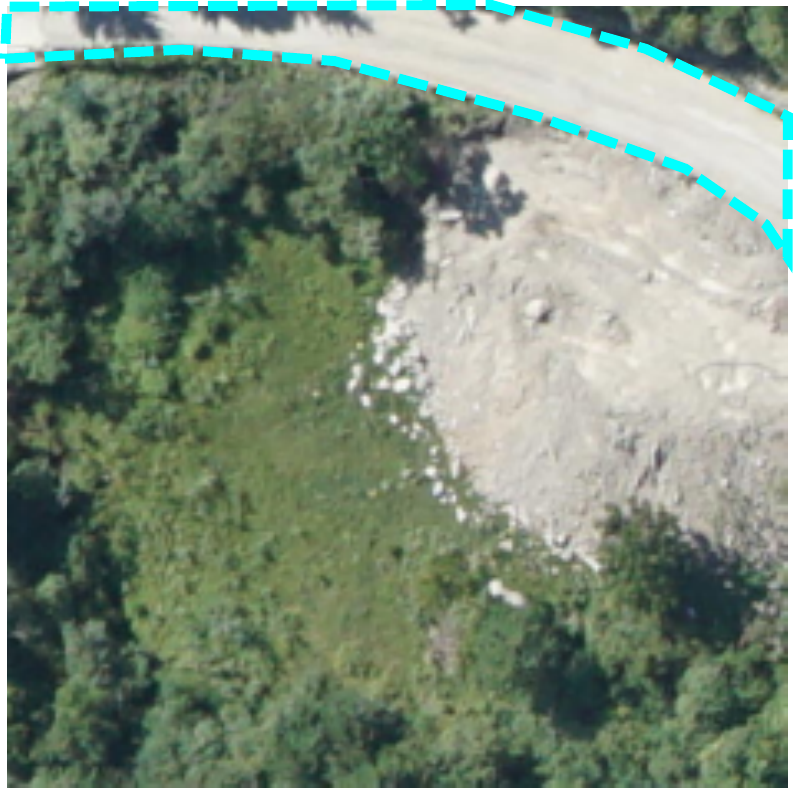}\vspace{0.3em}
    \includegraphics[width=0.1\linewidth]{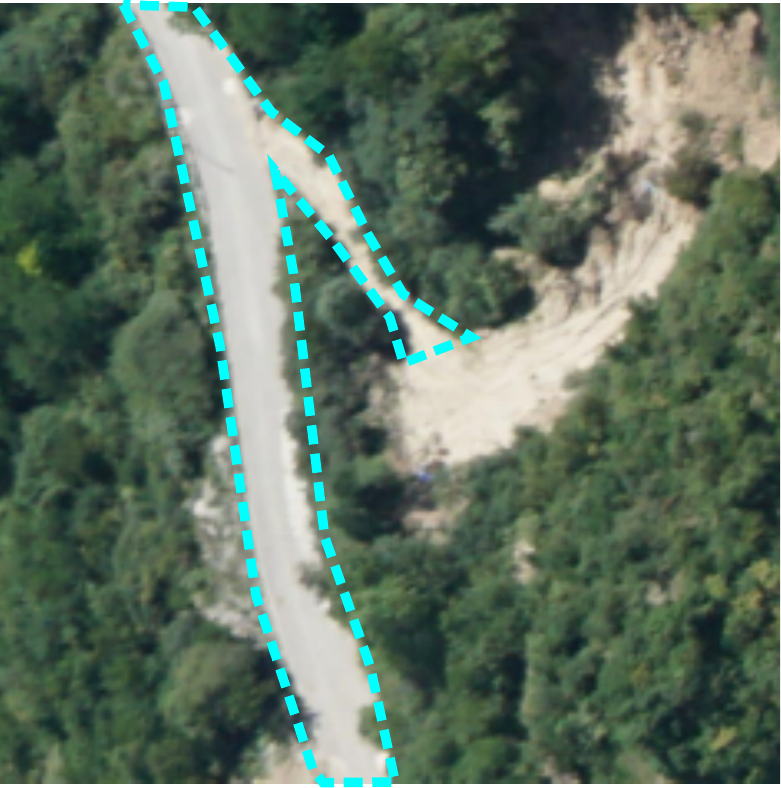}\vspace{0.3em}
    \includegraphics[width=0.1\linewidth]{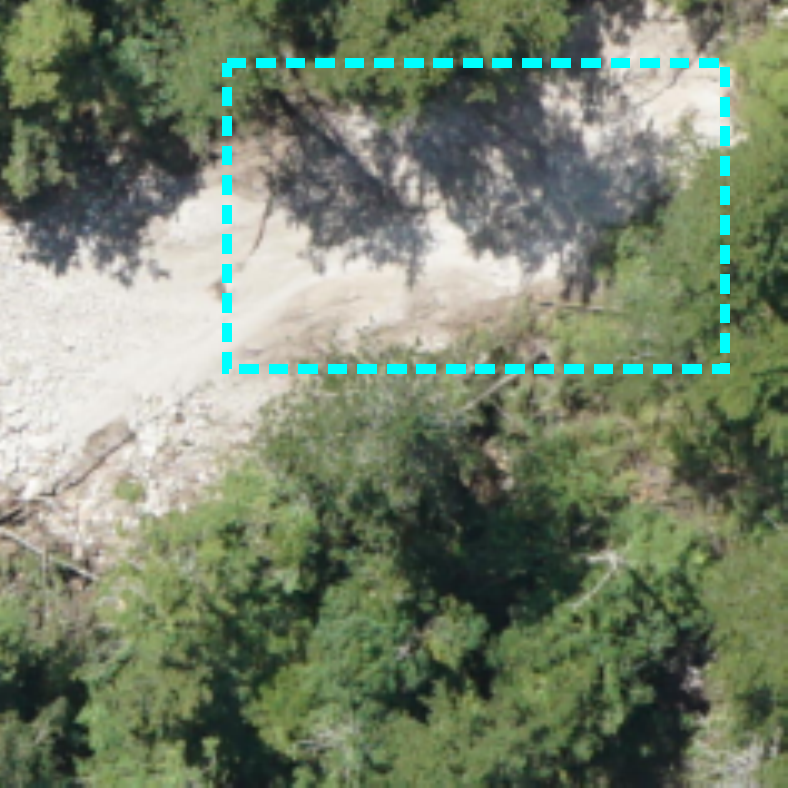}
  }
  \subcaptionbox{Label}[0.1\linewidth]{
    \includegraphics[width=0.1\linewidth]{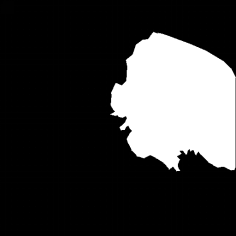}\vspace{0.3em}
    \includegraphics[width=0.1\linewidth]{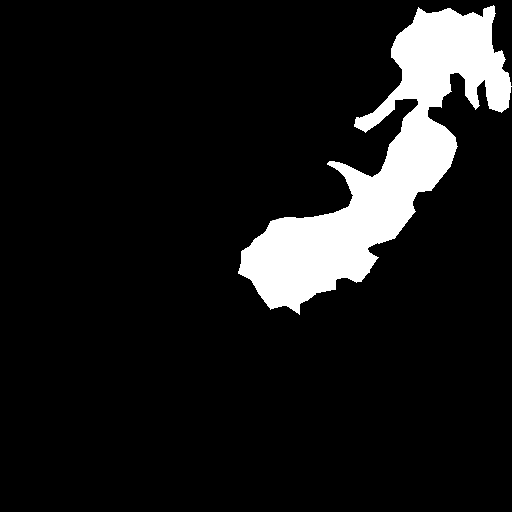}\vspace{0.3em}
    \includegraphics[width=0.1\linewidth]{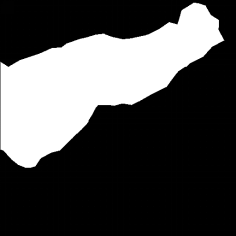}
  }
  \subcaptionbox{U-Net}[0.1\linewidth]{
    \includegraphics[width=0.1\linewidth]{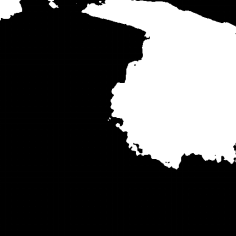}\vspace{0.3em}
    \includegraphics[width=0.1\linewidth]{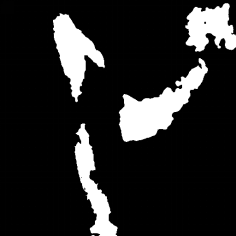}\vspace{0.3em}
    \includegraphics[width=0.1\linewidth]{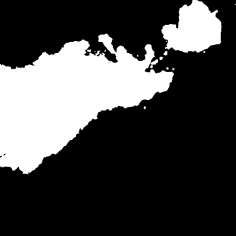}
  }
  \subcaptionbox{D-U-Net}[0.1\linewidth]{
    \includegraphics[width=0.1\linewidth]{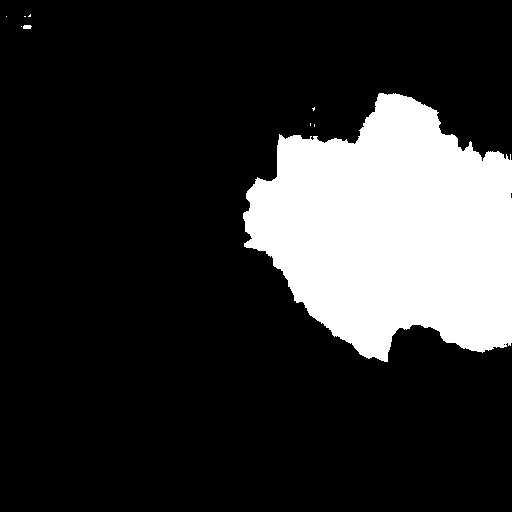}\vspace{0.3em}
    \includegraphics[width=0.1\linewidth]{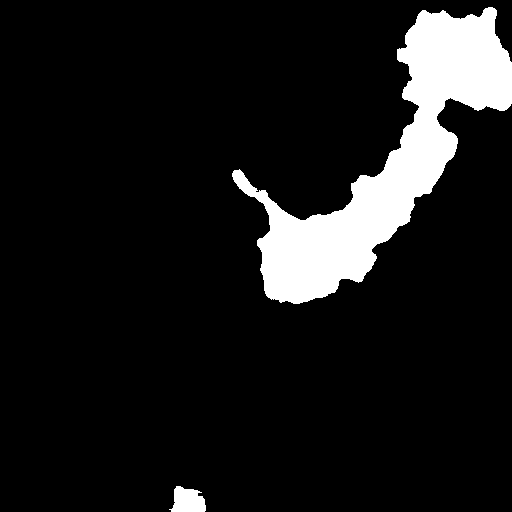}\vspace{0.3em}
    \includegraphics[width=0.1\linewidth]{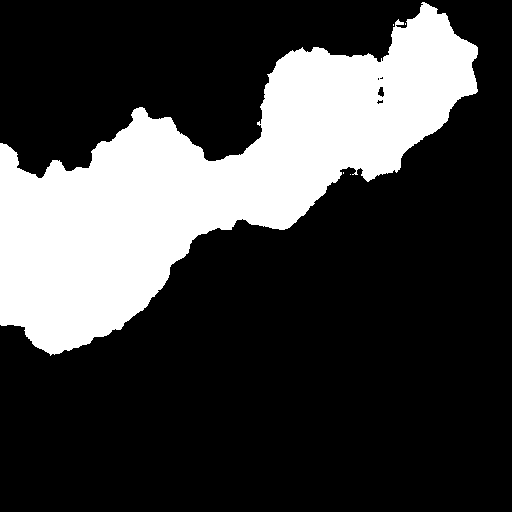}
  }
  \subcaptionbox{A-U-Net}[0.1\linewidth]{
    \includegraphics[width=0.1\linewidth]{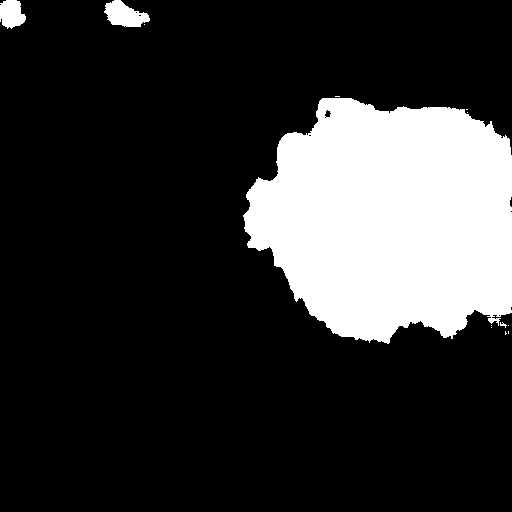}\vspace{0.3em}
    \includegraphics[width=0.1\linewidth]{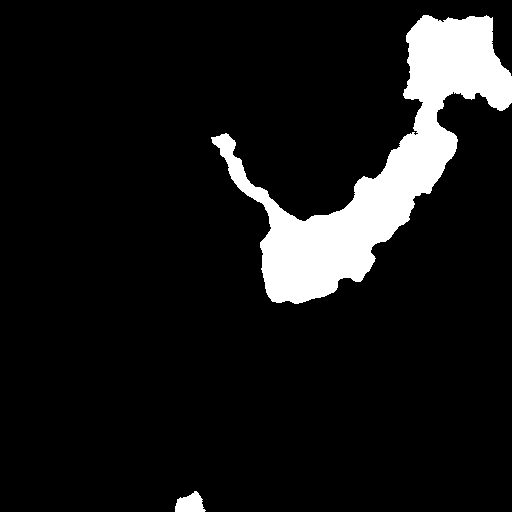}\vspace{0.3em}
    \includegraphics[width=0.1\linewidth]{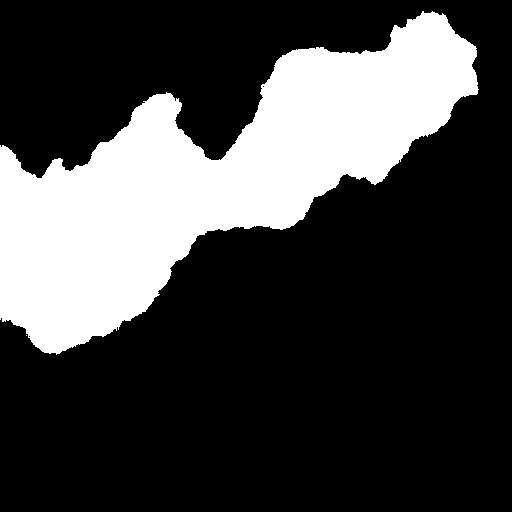}
  }
  \subcaptionbox{DA-U-Net}[0.1\linewidth]{
    \includegraphics[width=0.1\linewidth]{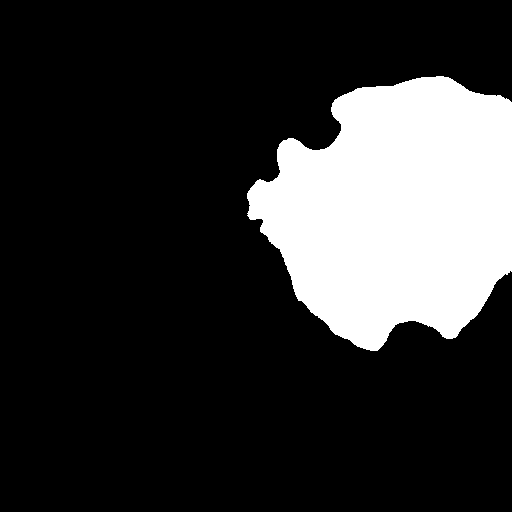}\vspace{0.3em}
    \includegraphics[width=0.1\linewidth]{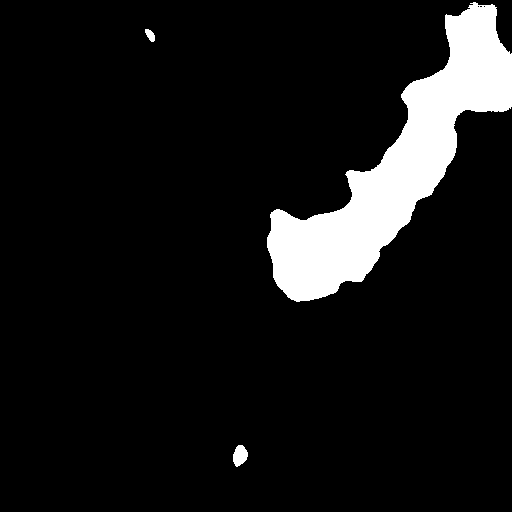}\vspace{0.3em}
    \includegraphics[width=0.1\linewidth]{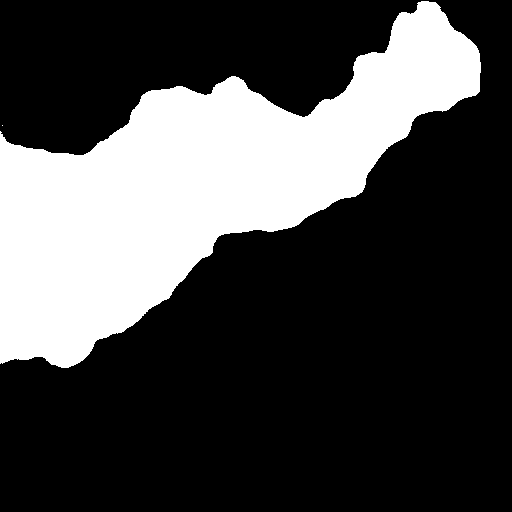}
  }
  \caption{Ablation studies of the DA-U-Net architecture. The highlighted are regions prone to errors.}
  \label{fig:ablation}
\end{figure*}

Fig. \ref{fig:ablation} compares different U-Net architectures in three confusing areas with roads. Both the D-U-Net and A-U-Net show improved detection of false landslide regions due to interference from roads, and attention-guided sampling shows better efficiencies than the dilation module and ASPP bottleneck. This is also consistent with the quantitative evaluations shown in Table \ref{tab:ablation}. Thus, in combination, these two modules demonstrated the best overall performances.

\begin{table}[h]
    \centering
    \caption{Ablation studies of different modules based on U-Net. The bold cell denotes the best performance.}
    \label{tab:ablation}
    \begin{tabular}{@{}lccc@{}}
    \toprule
    Method   & Attention Module & Dilated Convolution+ASPP & IOU   \\ \midrule
    U-Net    & $\times$                & $\times$                        & 48.18 \\
    D-U-Net & $\times$                & $\checkmark$                        & 54.61 \\
    A-U-Net & $\checkmark$                & $\times$                        & 52.44 \\
    DA-U-Net & $\checkmark$                & $\checkmark$                        & \textbf{59.41} \\ \bottomrule
    \end{tabular}
\end{table}

\section{Conclusion}
In this study we have proposed and developed an improved landslide inventory mapping methods that were based on augmenting the U-Net structure, \emph{i.e.} the dilated convolution \cite{chen2017deeplab} and attention-guided up-sampling \cite{chen2016attention,oktay2018attention}. As landslide or probable landslide regions generally co-exist with regions that also have similar spectral to landslides, contextual information should be considered to remove the feature-ambiguities produced by CNNs. The two modules are designed to fuse both local and non-local information to alleviate this issue. The two modules were used to simultaneously enlarge the receptive field of local convolution and preserve dense high-resolution feature maps. Future work may be devoted to the combined use of orthophotos and digital elevation models for more accurate and robust landslide mapping. In addition, pre-hazard susceptibility mapping of landslide-prone regions \cite{pradhan2010landslide} is also crucial for hazard mitigation. The code corresponding to this paper is made publicly available\footnote{https://github.com/saedrna/DA-U-Net}.

\bibliographystyle{IEEEtran}
\bibliography{landslide}
\end{document}